\documentclass[letterpaper, 10 pt, conference]{ieeeconf}  

\IEEEoverridecommandlockouts                              
\usepackage{amsmath,amssymb,amsfonts}
\usepackage{graphicx}
\usepackage{tikz}
\usepackage{pgfplots}
\pgfplotsset{compat=1.9}
\usepackage{pgfplotstable}
\usepgfplotslibrary{groupplots}
\usepackage[draft]{todonotes}
\usepackage{array} 
\usepackage{booktabs}
\usepackage[font=small,skip=0.3pt]{caption}
\usepackage{hyperref}

\DeclareMathOperator*{\argmin}{arg\,min}
\DeclareMathOperator{\Han}{\mathfrak{H}}

\newtheorem{assumption}{Assumption}

\newtheorem{lemma}{Lemma}
\newtheorem{corollary}[lemma]{Corollary}
\newtheorem{definition}{Definition}
\newtheorem{remark}{Remark}

\newcommand{\change}[1]{\textcolor{black}{#1}}

\title{\LARGE \bf
Physically Consistent Multiple-Step Data-Driven Predictions Using Physics-based Filters
}

\author{Yingzhao Lian$^*$, Jicheng Shi$^*$ and Colin N. Jones
\thanks{This work received support from the Swiss National Science Foundation (SNSF) under the NCCR Automation project, grant agreement 51NF40\_180545.The first two authors contributed equally.  \href{https://arxiv.org/abs/2303.09437}{Extended version: https://arxiv.org/abs/2303.09437} (corresponding author: Yingzhao Lian)}
\thanks{YL, JS and CNJ are with Automatic Laboratory, EPFL, 1015 Lausanne, Switzerland. {\tt\small $\{$yingzhao.lian, jicheng.shi, colin.jones$\}$@epfl.ch}}\vspace{-1em}}

\begin{document}

\maketitle

\begin{abstract}
Data-driven control can facilitate the rapid development of controllers, offering an alternative to conventional approaches. In order to maintain consistency between any known underlying physical laws and a data-driven decision-making process, preprocessing of raw data is necessary to account for measurement noise and any inconsistencies it may introduce. In this paper, we present a physics-based filter to achieve this and demonstrate its effectiveness through practical applications, using real-world datasets collected in a building on the École Polytechnique Fédérale de Lausanne (EPFL) campus. Two distinct use cases are explored: indoor temperature control and demand response bidding.
\end{abstract}

\section{Introduction}\label{sect:intro}
Data-driven control can improve the speed and quality of controller design and deployment via an end-to-end solution from I/O data to a functional controller. However, it is often crucial to ensure that the data-driven control should respect the known physical laws in order to make a meaningful decision. However, due to measurement noise present in the data, a direct use of raw data\footnote{Raw data in this work indicates the data without preprocessing.} may lead to incorrect conclusions or predictions. Such inconsistencies were spotted by~\cite{szegedy2013intriguing}, where minor perturbations in the input were shown to significantly deteriorate prediction accuracy~\cite{wiyatno2019adversarial}. 

The incorporation of physical laws in data-driven and machine learning methods has been an active area of research for decades. In fact, this idea has been used to solve partial differential equations since the 1990s~\cite{dissanayake1994neural}. The idea of incorporating a physical rule in a parametric model is referred to as "physics-guided" or "physics-informed" in the literature~\cite{cuomo2022scientific}. This can involve using the physical rule to define the loss function and to confine the model's parameters to a subset that is consistent with known physical rules. Researchers have applied this idea to various architectures, such as enforcing a positive correlation between indoor temperature and heating power consumption in neural networks~\cite{di2022physically}, and using a similar approach in linear parametric models~\cite{bunning2022physics}. While the aforementioned methods are important, preprocessing data can be a more direct approach to improve consistency. The methods falling in this category are highly related to robust optimization, where algorithms similar to scenario approaches have been successfully employed in natural language processing~\cite{brown2020language} and computer vision~\cite{xie2020self}.

In this work, we propose a physics-based filter that is tailored to data-driven control schemes based on Willems' fundamental lemma~\cite{willems2005note}. Willems' fundamental lemma offers a direct characterization of the system responses of linear-time-invariant (LTI) systems given an informative historical dataset. Such a characterization has been used in data-driven methods, and has been deployed in output prediction~\cite{markovsky2008data}, input reconstruction~\cite{turan2021data,shi2022data}, and in controller design~\cite{coulson2019data, de2019formulas, berberich2020data, lian2021adaptive,lian2021nonlinear}. \change{The main contribution lies in showing that some a priori knowledge can be integrated into Willems' fundamental lemma by robust optimization. The proposed scheme remains a non-parametric prediction structure, which differentiates it from other parametric schemes~\cite{di2022physically,bunning2022physics}.}

In order to present the proposed method with a more intuitive exposition, the idea presented in this paper will be motivated and related to building applications. In the following, the Willems' fundamental lemma and its corresponding prediction problem is reviewed in Section~\ref{sect:pre}, after which the physics-based filter is investigated in Section~\ref{sect:main}. The efficacy of the proposed scheme is validated on an indoor temperature control problem and a demand response bidding problem, with data collected from a building on the EPFL campus.

\noindent \textbf{Notation:} $I_n\in\mathbb{R}^{n\times n}$ denotes a $n$-by-$n$ identity matrix, similarly, we denote the zero matrix by $\mathbf{O}$. $\mathbf{0}$ and $\mathbf{1}$ respectively denotes a zero vector and a one vector. $\mathrm{blkdiag}(A_1,\dots,A_n)$ generates a block-diagonal matrix whose diagonal blocks are $A_1,\dots,A_n$ accordingly. $x: = \{x_i\}_{i=1}^T$ denotes a sequence of size $T$ indexed by $i$. $x_i$ denotes the measurement of $x$ at time $i$, and $x_{1:L}:=[x_1^\top,x_2^\top\dots x_L^\top]^\top$ denotes a concatenated sequence of $x_i$ ranging from $x_1$ to $x_L$, and we drop the index to improve clarity if the intention is clear from the context.

\vspace{-.3em}
\section{Preliminaries}\label{sect:pre}
\begin{definition}
A Hankel matrix of depth $L$ associated with a vector-valued signal sequence $s:=\{s_i\}_{i=1}^T,\;s_i\in\mathbb{R}^{n_s}$ is
\begin{align*}
    \Han_L(s):=
    \begin{bmatrix}
    s_1 & s_2&\dots&s_{T-L+1}\\
    s_2 & s_3&\dots&s_{T-L+2}\\
    \vdots &\vdots&&\vdots\\
    s_{L} & s_{L+1}&\dots&s_T
    \end{bmatrix}\;.
\end{align*}
\end{definition}

A linear time-invariant (LTI) system is defined by $ x_{i+1} = Ax_i+Bu_i\;,\;y_i = Cx_i+Du_i$, dubbed $\mathfrak{B}(A,B,C,D)$. Its order is $n_x$ with $n_u,\enspace n_y$ denoting its input and output dimensions respectively. An $L$-step trajectory generated by this system is $\begin{bmatrix}u_{1:L} & y_{1:L}\end{bmatrix}:=\begin{bmatrix}
         u_1^\top&\dots & u_{L}^\top & y_{1}^\top &\dots&y_{L}^\top
    \end{bmatrix}^\top$.
The set of all possible $L$-step trajectories generated by $\mathfrak{B}(A,B,C,D)$ is denoted by $\mathfrak{B}_L(A,B,C,D)$. 
For the sake of consistency, a datapoint coming from the historical dataset is marked by boldface subscript $_\textbf{d}$. Given a sequence of input-output measurements $\{u_{\textbf{d},i},y_{\textbf{d},i}\}_{i}$, we call the input sequence persistently exciting of order $L$ if $\Han_L(u_\textbf{d})$ is full row rank. By building the following stacked Hankel matrix $\Han_L(u_\textbf{d}, y_\textbf{d}):=\begin{bmatrix}
    \Han_{L}(u_{\textbf{d}})^\top &\Han_L(y_{\textbf{d}})^\top
    \end{bmatrix}^\top$,
we state \textbf{Willems' Fundamental Lemma} as
\begin{lemma}\label{lem:funda}\cite[Theorem 1]{willems2005note}
Consider a controllable linear system and assume $\{u_\textbf{d}\}_{i=1}^T$ is persistently exciting of order $L+ n_x$. The condition $\text{colspan}(\Han_L(u_\textbf{d},y_\textbf{d}))=\mathfrak{B}_L(A,B,C,D)$ holds.
\end{lemma}

For the sake of consistency, $L$ is reserved for the length of the system responses. A data-driven control scheme has been proposed in~\cite{coulson2019data,markovsky2007linear}, where Lemma~\ref{lem:funda} generates a trajectory prediction. Before introducing the prediction, we state the following assumption to simplify the presentation 
of this paper:
\begin{assumption}\label{ass:noisy_output}
The output measurements $y$ are contaminated by measurement noise, the input measurements $u$ are exact.
\end{assumption}
It is possible to consider noisy input measurements; please refer to~\cite{lian2021adaptive} for more details. Under Assumption~\ref{ass:noisy_output}, the trajectory prediction problem is defined by:
\begin{subequations}\label{eqn:pred}
\begin{align}
    &\;y_{pred}(u_{pred}) = \Han_{L,pred}(y_\textbf{d})g \\
        g&\in\argmin_{g_l,\sigma_l} \frac{1}{2}\lVert\sigma_l\rVert^2+\frac{1}{2}g_l^\top \mathcal{E}_g g_l\label{eqn:pred_obj}\\
        &\quad\quad\quad\text{s.t.}\;\begin{bmatrix}
        \Han_{L,init}(y_\textbf{d})\\\Han_{L,init}(u_\textbf{d})\\\Han_{L,pred}(u_\textbf{d})
        \end{bmatrix}g_l=\begin{bmatrix}
        y_{init}+\sigma_l\\u_{init}\\u_{pred}
    \end{bmatrix}\nonumber\;,
\end{align}
\end{subequations}
where $\mathcal{E}_g$ is a user-defined \change{positive definite} penalty and $u_{init}, y_{init}$ are $t_{init}$-step sequences of the measured inputs and outputs preceding the current point in time. Accordingly, $u_{pred},y_{pred}$ are the corresponding $n_h$-step predictive sequences viewed from the current time step. The matrix $\Han_L(y_\textbf{d})$ is split into two sub-Hankel matrices:
\begin{align*}
    \Han_L(y_\textbf{d}) = \begin{bmatrix}
        \Han_{L,init}(y_\textbf{d})\\\Han_{L,pred}(y_\textbf{d})
    \end{bmatrix}\;.
\end{align*}
The matrix $\Han_{L,init}(y_\textbf{d})$ is of depth $t_{init}$ and the depth of $\Han_{L,pred}(y_\textbf{d})$ is the prediction horizon $n_h$ such that $t_{init}+n_h = L$. The matrices $\Han_{L,init}(u_\textbf{d})$, $\Han_{L,pred}(u_\textbf{d})$ are defined similarly. The choice of $t_{init}$ is made to ensure a unique estimation of the initial state; please refer to~\cite{markovsky2008data} for more details. This prediction problem~\eqref{eqn:pred} predicts $n_h$-step output trajectory $y_{pred}$ for any given predictive input sequence $u_{pred}$, whose objective in~\eqref{eqn:pred_obj} minimizes a Wasserstein distance upper bound; the interested readers are referred to~\cite{lian2021adaptive} for more details. Further, recalling the conditions of Willems' fundamental Lemma~\ref{lem:funda}, this prediction problem requires the following assumption:
\begin{assumption}\label{ass:PE}
$u_{\mathbf{d}}$ is persistently exciting of order $L+n_x$.
\end{assumption}

\vspace{-.3em}
\section{Main Results}\label{sect:main}
\subsection{Physics-based Filter} \label{sect:main_filter}
As discussed in Section~\ref{sect:intro}, the measurement noise presented in $\{y_\textbf{d}\}$ may lead to inconsistent output predictions in~\eqref{eqn:pred}. Hence, the data preprocessing scheme should modify the data $\{y_\mathbf{d}\}$ such that the prediction generated by~\eqref{eqn:pred} is consistent with some prior physical rules. \change{Here we focus on the following two rules from building control applications:}
\begin{itemize}
    \item \textbf{Temperature consistency:} The indoor temperature is positively correlated with the power consumption of the heating, cooling and ventilation (HVAC) system. More specifically, if the room is heated by control input $u_{pred}$, the predicted indoor temperature must be higher than the predicted temperature that is controlled by $u_{pred}=\mathbf{0}$.
    \item \textbf{Bidding consistency:} Demand response (DR) is a method of managing power demand on the consumption side~\cite{vazquez2010energy}. If a building is to provide, for example, secondary frequency control services, it tracks an area generation control (AGC) signal provided by the transmission system operator (TSO), while maintaining indoor comfort. \change{Intuitively speaking, the TSO manipulates the building as a slow but large-capacity ``battery", and as a result, a higher/lower power consumption than its nominal value relatively ``charge/discharge" the ``battery". The ``capacity" of the battery is accordingly central to its flexibility in the context of DR, which is reflected by the accumulative indoor temperature relative to that operated by the nominal power consumption. Note that the absolute power consumption is still non-negative. The minimal physical rule to ensure a reasonable bidding proposal is therefore the positive correlation between the accumulated indoor temperature and power consumption (i.e.$\sum_i y_{pred,i} \ge 0,\forall\; u_{pred}\ge0$).}
\end{itemize}

Drawing inspiration from the discussion above, we can identify the essential components required to define a physics-based filter:
\begin{itemize}
    \item \change{The convex set $\mathcal{Y}$ of trajectories that is aligned with the physical rule}, and $y_{pred}$ is consistent if $y_{pred}\in\mathcal{Y}$.
    \item The set of control inputs $\mathcal{U}$ and initial conditions $\mathcal{U}_{init},\;\mathcal{Y}_{init}$ where the physical rule is imposed. 
\end{itemize}
Recall the aforementioned examples, their mathematical components are defined by \change{(see Remark~\ref{rmk:rule_temp} for more details)}:
\begin{itemize}
    \item \textbf{Temperature consistency:} 
    \begingroup\makeatletter\def\f@size{9.5}\check@mathfonts
    \begin{align}\label{eqn:rule_temp}
           \hspace{-2em} \mathcal{Y}=\left\{y\middle|y\geq \mathbf{0}\right\},\;\mathcal{Y}_{init}=\textbf{0},\;
            \mathcal{U}=\left\{u|u\geq \mathbf{0}\right\},\; \mathcal{U}_{init}=\textbf{0}
    \end{align}
    \endgroup
    \item \textbf{Bidding consistency:}
    \begingroup\makeatletter\def\f@size{9.5}\check@mathfonts
    \begin{align}\label{eqn:rule_bidding}
            \hspace{-2em}\mathcal{Y}=\left\{y\middle|\mathbf{1}^\top y\geq 0\right\},\mathcal{Y}_{init}=\textbf{0},
            \mathcal{U}=\left\{u|u\geq \mathbf{0}\right\},\mathcal{U}_{init}=\textbf{0}
    \end{align}
    \endgroup
\end{itemize}
Accordingly, the physics-based filter is defined by the following robust optimization problem:
\begin{subequations}\label{eqn:filter}
\begin{align}
        \min\limits_{\tilde{y}_\mathbf{d}} \;&\; \lVert \Tilde{y}_{\mathbf{d}} - y_\mathbf{d}\rVert\label{eqn:filter_obj}
\end{align}
subject to: $\forall\; u_{pred}\in \mathcal{U},\;u_{init}\in\mathcal{U}_{init},\;y_{init}\in\mathcal{Y}_{init}$
\begin{align}
    \hspace{-2.8em}y_{pred} = \Han_{L,pred}(\Tilde{y}_{\mathbf{d}})g \in\mathcal{Y}\label{eqn:filter_ypred}
\end{align}
\begin{align}
        g&\in\argmin_{g_l,\sigma_l} \frac{1}{2}\lVert\sigma_l\rVert^2+\frac{1}{2}g_l^\top \mathcal{E}_g g_l\nonumber\\
        &\quad\quad\text{s.t.}\;\begin{bmatrix}
        \Han_{L,init}(\Tilde{y}_{\mathbf{d}})\\\Han_{L,init}(u_\textbf{d})\\\Han_{L,pred}(u_\textbf{d})
        \end{bmatrix}g_l=\begin{bmatrix}
        y_{init}+\sigma_l\\u_{init}\\u_{pred}
    \end{bmatrix}\label{eqn:filter_lower_cons}.
\end{align}
\end{subequations}
This is a bi-level robust optimization problem, which minimizes the perturbation of the offline dataset $\{y_\mathbf{d}\}$. Particularly, the post-processed output data $\{\Tilde{y}_{\mathbf{d}}\}$ will replace the raw data $\{y_\mathbf{d}\}$ in the definition of the prediction problem. The robust constraint enforces that, for any possible predictive input sequence $u_{pred}\in\mathcal{U}$, the corresponding output sequence $y_{pred}$ should be consistent with the physical rule in~\eqref{eqn:filter_ypred}. In the next section, we will show how to convert this problem~\eqref{eqn:filter} into a numerically tractable form. 
\subsection{Single-level Reformulation}
Regardless of the physical rule $\mathcal{Y}$, solving a bi-level optimization can be non-trivial. However, in this case, the physics-based filter~\eqref{eqn:filter} can be reformulated into a single-level optimization problem:
\begin{lemma}\label{lem:reform_kkt}
The following single-level problem is equivalent to the bi-level problem~\eqref{eqn:filter}:
\begin{subequations}\label{eqn:filter_single}
\begin{align}
        \min\limits_{\tilde{y}_\mathbf{d}} \;&\; \lVert \Tilde{y}_{\mathbf{d}} - y_\mathbf{d}\rVert\\
    \text{s.t.} &\; \forall\; u_{pred}\in \mathcal{U},\;u_{init}\in\mathcal{U}_{init},\;y_{init}\in\mathcal{Y}_{init}\notag\\
    &\; \;y_{pred} = \Han_{L,pred}(\Tilde{y}_{\mathbf{d}})g\in\mathcal{Y}\notag\\
        &\;M(\tilde{y}_\mathbf{d})\begin{bmatrix}g\\\kappa(u_{pred})\end{bmatrix} = \begin{bmatrix}\Han_{L,init}(\tilde{y}_\textbf{d})^\top y_{init}\\u_{init}\\u_{pred}\end{bmatrix}\;\label{eqn:filter_kkt},
\end{align}
\end{subequations}
where $\kappa(u_{pred})$ is the dual variable of~\eqref{eqn:filter_lower_cons} and
\begingroup\makeatletter\def\f@size{9.5}\check@mathfonts
\begin{align}
    M(\tilde{y}_\mathbf{d}) := \begin{bmatrix}
        \Han_{L,init}(\tilde{y}_\textbf{d})^\top\Han_{L,init}(\tilde{y}_\textbf{d})+\mathcal{E}_g &\Han_L(u_\textbf{d})^\top\\
        \Han_L(u_\textbf{d})&\textbf{O}
        \end{bmatrix}.\label{eqn:filter_kkt_mat}
\end{align}
\endgroup
\end{lemma}
\vspace{.5em}

\begin{proof}
Note that the lower level problem in~\eqref{eqn:filter} is strongly convex, it is therefore equivalent to its KKT system~\cite[Chapter 4]{dempe2002foundations}. By replacing $\sigma_l$ by $\Han_{L,init}(y_\textbf{d})g_l-y_{init}$, the Lagrangian of the lower level problem is
\begin{align*}
    \mathcal{L}(g) = \frac{1}{2}\lVert \Han_{L,init}(y_\textbf{d})g&-y_{init}\rVert^2+\frac{1}{2}g^\top\mathcal{E}_g g\\
    &+\kappa^\top (\Han_L(u_\textbf{d})g-\begin{bmatrix}
    u_{init}\\u_{pred}
    \end{bmatrix})\;,
\end{align*}
where $\kappa(u_{pred})$ is the dual variable of the equality constraint. Hence, we have the stationary condition of the KKT system:
\begin{align*}
    \frac{\partial \mathcal{L}(g)}{\partial g}^\top = (\Han_{L,init}(y_\textbf{d})^\top&\Han_{L,init}(y_\textbf{d})+\mathcal{E}_g)g +\Han_L(u_\textbf{d})^\top\kappa\\&-\Han_{L,init}(y_\textbf{d})^\top y_{init} = 0\;.
\end{align*}
By recalling the primal feasibility condition
\begin{align*}
    \Han_L(u_\textbf{d})g = \begin{bmatrix}
    u_{init}^\top&u_{pred}^\top
    \end{bmatrix}^\top\;,
\end{align*}
we get the robust KKT matrix $M$ in~\eqref{eqn:filter_kkt_mat}.
\end{proof}

\begin{remark}\label{rmk:rule_temp}
\change{The physical rules~\eqref{eqn:rule_temp} and~\eqref{eqn:rule_bidding} are defined on the transient response, which is linear with respect to $u_{pred}$ in LTI systems. In the rule of temperature consistency~\eqref{eqn:rule_temp}, our a priori knowledge requires that if $u_{pred}\geq\tilde{u}_{pred}$, their corresponding transient responses satisfy $y_{pred}\geq\tilde{y}_{pred}$. By the superposition property, $y_{pred}-\tilde{y}_{pred}$ is the transient response of $u_{pred}-\tilde{u}_{pred}$, which summarizes the rule in~\eqref{eqn:rule_temp}.}
\end{remark}
\begin{remark}
\change{Assumption~\ref{ass:PE} is not strong in building applications, the stochastic property of the process noise (e.g. solar radiation and outdoor weather) will cause random fluctuation in the closed-loop input trajectory, and the persistent excitation condition is in turn satisfied.}
\end{remark}

\subsection{Affine Physical Rules}\label{sect:affine_rule}
Recall the physical rules mentioned in Section~\ref{sect:main_filter}, we are particularly interested in affine physical rules, i.e. $\mathcal{Y}= \{y|H_{y,pred}y\leq h_{y,pred}\},\;\mathcal{Y}_{init}= \{y|H_{y,init}y\leq h_{y,init}\},\;\mathcal{U}_{init}= \{u|H_{u,init}u\leq h_{u,init}\}$ and $\mathcal{U}=\{u|H_{u,pred} u\leq h_{u,pred}\}$. A tractable reformulation for the affine physical rule is stated in the following corollary.
\begin{corollary}\label{cor:affine_rule}
Consider an affine physical rule. The solution to the physics-based filter~\eqref{eqn:filter_single} is equivalent to the solution to the following problem:
\begingroup\makeatletter\def\f@size{9.6}\check@mathfonts
\begin{align}
    &\min\limits_{\nu\geq \mathbf{O},\;\tilde{y}_\mathbf{d}}\;\; \lVert \Tilde{y}_{\mathbf{d}} - y_\mathbf{d}\rVert\\
    &\text{s.t.}\; h_{y,pred}\geq  h_\mathrm{aug}(\tilde{y}_\mathbf{d})^\top\nu,\;M_{\mathrm{aug}}(\tilde{y}_\mathbf{d})^\top\lambda +H_\mathrm{aug}\nu = H_{\mathrm{obj}}\;\notag.
\end{align}
where $x=\begin{bmatrix}
     g^\top & \kappa^\top &y_{init}^\top&u_{init}^\top&u_{pred}^\top
\end{bmatrix}^\top$,
\begin{align*}
    H_\mathrm{obj}(\tilde{y}_\mathbf{d}):=&\begin{bmatrix}
         H_{y,pred}\Han_{L,pred}(\tilde{y}_\mathbf{d})&\change{\mathbf{O}}
    \end{bmatrix}\\
    M_\mathrm{aug}(\tilde{y}_\mathbf{d}):=& \begin{bmatrix}
         M(\tilde{y}_\mathbf{d}) &\mathrm{blkdiag}(-\Han_{L,init}(\tilde{y}_\textbf{d})^\top, -I, -I)
    \end{bmatrix}\;,\\
    H_\mathrm{aug} :=& \begin{bmatrix}
         \mathbf{O}&\mathrm{blkdiag}(H_{y,init},H_{u,init},H_{u,pred})
    \end{bmatrix}\;,
\end{align*}
\begin{align*}
    h_\mathrm{aug}:=&\begin{bmatrix}
         h_{y,init}^\top&h_{u,init}^\top&h_{u,pred}^\top
    \end{bmatrix}^\top\;.
\end{align*}
\endgroup
\end{corollary}
\vspace{.1em}
\begin{proof}
The physics-based filter under an affine physical rule is defined by following robust optimization problem:
\begin{align*}
            \min\limits_{\tilde{y}_\mathbf{d}} \;&\; \lVert \Tilde{y}_{\mathbf{d}} - y_\mathbf{d}\rVert\\
    \text{s.t.} &\;h_{y,pred}\geq \max\limits_{\substack{u_{init},y_{init}\\u_{pred}}} H_{y,pred}\Han_{L,pred}(\Tilde{y}_{\mathbf{d}})g\\
        &\quad\;\text{s.t}\begin{cases} H_{u,pred}u_{pred}\leq h_{u,pred},
    H_{u,init}u_{init}\leq h_{u,init}\\
     H_{y,init}y_{init}\leq h_{y,init}\\
     M(\tilde{y}_\mathbf{d})\begin{bmatrix}g\\\kappa\end{bmatrix} = \begin{bmatrix}\Han_{L,init}(\tilde{y}_\textbf{d})^\top y_{init}\\u_{init}\\u_{pred}\end{bmatrix},
    \end{cases}
\end{align*}
which can be reformulated into the standard form of LP:
\begin{align*}
    \min\limits_{\tilde{y}_\mathbf{d}}\;&\; \lVert \Tilde{y}_{\mathbf{d}} - y_\mathbf{d}\rVert\\
    \text{s.t.}\;& h_{y,pred}\geq \max_x H_{\mathrm{obj}}(\tilde{y}_\mathbf{d})x\\
    &\quad\; \text{s.t.}\; M_\mathrm{aug}(\tilde{y}_\mathbf{d})x= \mathbf{0},\;
    H_\mathrm{aug}x\leq h_\mathrm{aug}\;.
\end{align*}
By duality of LP~\cite{ben2009robust}, the constraint is reformulated to
\begin{align*}
    \min\limits_{\tilde{y}_\mathbf{d}}\;&\; \lVert \Tilde{y}_{\mathbf{d}} - y_\mathbf{d}\rVert\\
    \text{s.t.}\;& h_{y,pred}\geq \min_{\lambda,\nu}\; h_\mathrm{aug}(\tilde{y}_\mathbf{d})^\top\nu\\
    &\text{s.t.}\;
    M_{\mathrm{aug}}(\tilde{y}_\mathbf{d})^\top\lambda +H_\mathrm{aug}\nu = H_{\mathrm{obj}},\;
    \nu\geq \mathbf{O}\;.
\end{align*}
This is sufficient to summarize the proof.
\end{proof}
\begin{remark}
As an optimization problem still must be solved in the proposed scheme, one may question its benefit. We summarize the scenarios in which proposed scheme is advantageous to a parametric system identification approach:
\begin{itemize}
    \item When the physical rules are defined based on the I/O sequence, such as the \change{passivity~\cite{van2006port}, the independence/causality between I/O ports~\cite{fattahi2019graphical} and positive/negative correlation (e.g. rules in this work)}, using the proposed scheme is more intuitive without converting the physical rule to its parametric correspondence.
    \item If a physical rule is defined by a multi-step I/O sequence, a parametric model may involve high-order polynomials on its parameters that is not desirable for numerical solvers. Consider a one dimensional case with $y_{i+1}=ay_i+bu_i$; the parametric form of the bidding consistency is defined by a high-order polynomial $\sum_{i=0}^{n_h-1}a^ib\geq 0$. Instead, the dual solved in the proposed method remains bilinear (see Section~\ref{sect:num}).
\end{itemize}
\end{remark}

\subsection{Numerical Details}\label{sect:num}
The reformulated single-level problem~\eqref{eqn:filter_single} is still a non-convex optimization due to the nonlinear equality constraint~\eqref{eqn:filter_kkt}, where the quadratic term $\Han_{L,init}^\top(\tilde{y}_\textbf{d})\Han_{L,init}(\tilde{y}_\textbf{d})$ in the matrix $M(\tilde{y}_\textbf{d})$ is numerically less desirable to most optimization solvers. In order to improve the numerical performance, we suggest reformulating the problem~\eqref{eqn:filter_single} as
\vspace{-1em}
\begin{subequations}\label{eqn:filter_single_reform}
\begin{align}
        \min\limits_{\tilde{y}_\mathbf{d}} \;&\; \lVert \Tilde{y}_{\mathbf{d}} - y_\mathbf{d}\rVert\\
    \text{s.t.} &\; \forall\; u_{pred}\in \mathcal{U},\;u_{init}\in\mathcal{U}_{init},\;y_{init}\in\mathcal{Y}_{init}\notag\\
    &\; \;y_{pred} = \Han_{L,pred}(\Tilde{y}_{\mathbf{d}})g\in\mathcal{Y}\notag
\end{align}
\begin{align}
        &\;M_{\mathrm{sch,1}}(\tilde{y}_\mathbf{d})\begin{bmatrix}\sigma\\g\\\kappa\end{bmatrix} = \begin{bmatrix}y_{init}^\top&\mathbf{0}^\top&u_{init}^\top&u_{pred}^\top\end{bmatrix}^\top\label{eqn:filter_kkt_reform}
\end{align}
\end{subequations}
where 
\begin{align*}
    M_{\mathrm{sch,1}}(\tilde{y}_\mathbf{d}) &:= \begin{bmatrix} -I &\Han_{L,init}(\tilde{y}_\textbf{d})&\mathbf{O}\\
        \Han_{L,init}(\tilde{y}_\textbf{d})^\top&\mathcal{E}_g &\Han_L(u_\textbf{d})^\top\\
        \mathbf{O}&\Han_L(u_\textbf{d})&\textbf{O}
        \end{bmatrix}\;.
\end{align*}
The equivalence between~\eqref{eqn:filter_kkt} and~\eqref{eqn:filter_kkt_reform} follows the Schur complement (i.e. inverse Gaussian elimination)~\cite{zhang2006schur}. The benefit of using~\eqref{eqn:filter_kkt_reform} instead of~\eqref{eqn:filter_kkt} is that, the right-hand side of~\eqref{eqn:filter_kkt_reform} is independent of $\tilde{y}_{\mathbf{d}}$ and the left-hand side is linear with respect to $\Han_{L,init}^\top(\tilde{y}_\textbf{d})$ instead of quadratic. Even though problem~\eqref{eqn:filter_single_reform} is still non-convex \change{due to the bilinearity induced by the multiplication between $\Han_{L,init}(\tilde{y}_\mathbf{d})$ and $g$ in~\eqref{eqn:filter_kkt_reform}, there exist more efficient and reliable numerical optimization algorithms tailored for bilinear problems,} such as the McCormick envelope~\cite{mccormick1976computability} implemented in a recent release of \textsc{Gurobi 9.0}~\cite{achterberg2019s}. Based on our numerical experiment, this reformulation can roughly gain 50\% acceleration in the solution time with the same initialization.

In addition to the benefits in numerical efficiency, the reformulation given in~\eqref{eqn:filter_kkt_reform} is particularly valuable when using the horizon splitting technique. As reported in~\cite{o2022data}, horizon splitting can improve long-term prediction accuracy, which is central to the bidding problem in DR. Under a horizon splitting scheme, the predictor given by equation~\eqref{eqn:pred} is recursively called to generate a long prediction trajectory by concatenation. Without loss of generality, we explain it by a special case where $t_{init}=n_h$, and a prediction trajectory of $2n_h$-steps is generated. This prediction is obtained by solving the following optimization problem
\begingroup\makeatletter\def\f@size{9.6}\check@mathfonts
\begin{subequations}\label{eqn:pred_split}
\begin{align}
    &\;y_{pred,1} = \Han_{L,pred}(y_\textbf{d})g_1 \notag\\
        g_1&\in\argmin_{g_l,\sigma_l} \frac{1}{2}\lVert\sigma_l\rVert^2+\frac{1}{2}g_l^\top \mathcal{E}_g g_l\notag\\
        &\quad\quad\text{s.t.}\;\begin{bmatrix}
        \Han_{L,init}(y_\textbf{d})\\\Han_{L,init}(u_\textbf{d})\\\Han_{L,pred}(u_\textbf{d})
        \end{bmatrix}g_l=\begin{bmatrix}
        y_{init}+\sigma_l\\u_{init}\\u_{pred,1}
    \end{bmatrix}\label{eqn:pred_split_cons1}\\
    &\;y_{pred,2} = \Han_{L,pred}(y_\textbf{d})g_2 \notag
\end{align}
\begin{align}
        g_2&\in\argmin_{g_l,\sigma_l} \frac{1}{2}\lVert\sigma_l\rVert^2+\frac{1}{2}g_l^\top \mathcal{E}_g g_l\notag\\
        &\quad\text{s.t.}\;\begin{bmatrix}
        \Han_{L,init}(y_\textbf{d})\\\Han_{L,init}(u_\textbf{d})\\\Han_{L,pred}(u_\textbf{d})
        \end{bmatrix}g_l=\begin{bmatrix}
        y_{pred,1}+\sigma_l\\u_{pred,1}\\u_{pred,2}
    \end{bmatrix}\label{eqn:pred_split_cons2}
\end{align}
\end{subequations}
\endgroup
where the predictive input sequence $u_{pred}$ of length $2n_h$ is partitioned into two $n_h$-step sequences, i.e. $u_{pred}=\begin{bmatrix}
     u_{pred,1}^\top&u_{pred,2}^\top
\end{bmatrix}^\top$. Similarly, we have $y_{pred,1}$ and $y_{pred,2}$. 
The predictive component in~\eqref{eqn:pred_split_cons1} composes the initialization component in~\eqref{eqn:pred_split_cons2}. 
The formulation~\eqref{eqn:filter_kkt_reform} plays a crucial role in enabling numerically efficient implementation. By utilizing the single level-reformulation provided in Lemma~\ref{lem:reform_kkt}, the resulting physics-based filter is 
defined as follows:
\begingroup\makeatletter\def\f@size{9.6}\check@mathfonts
\begin{subequations}
\begin{align*}
    \min\limits_{\tilde{y}_\mathbf{d}} \;&\; \lVert \Tilde{y}_{\mathbf{d}} - y_\mathbf{d}\rVert\notag\\
    \text{s.t} &\; \forall\; u_{pred}\in \mathcal{U},\;u_{init}\in\mathcal{U}_{init},\;y_{init}\in\mathcal{Y}_{init}\notag\\
&\; \;y_{pred}(u_{pred}) =\Han_{L,pred}(\Tilde{y}_{\mathbf{d}})
    \begin{bmatrix}
           g_1^\top&
           g_2^\top
    \end{bmatrix}^\top\in\mathcal{Y}\notag
\end{align*}
\begin{align*}
&\;M_{\mathrm{sch,2}}(\tilde{y}_\mathbf{d})\begin{bmatrix}\sigma_1\\g_1\\\kappa_1\\\sigma_2\\g_2\\\kappa_2\end{bmatrix} = \begin{bmatrix}y_{init}\\\mathbf{0}\\u_{init}\\u_{pred,1}\\\mathbf{0}\\\mathbf{0}\\u_{pred,1}\\u_{pred,2}\end{bmatrix}
\end{align*}
\end{subequations}
\endgroup
where 
\begin{align*}
    M_{\mathrm{sch,2}}(\tilde{y}_\mathbf{d}) &:= \begin{bmatrix} M_{\mathrm{sch,1}}(\tilde{y}_\mathbf{d})&\mathbf{O}\\
    \begin{bmatrix}
           \mathbf{O}&-\Han_{L,pred}(\tilde{y}_\mathbf{d})&\mathbf{O}\\
           \mathbf{O}&\mathbf{O}&\mathbf{O}
        \end{bmatrix}&M_{\mathrm{sch,1}}(\tilde{y}_\mathbf{d})
        \end{bmatrix}\;.
\end{align*}
Although the data-driven predictor is recursively called twice, the resulting optimization problem remains bilinear. In general, by applying the inverse Schur complement technique in~\eqref{eqn:filter_kkt_reform}, the physics-based filter remains bilinear regardless of the number of segments that the predictive trajectory is split into. \change{It is worth mentioning that the reformulation suggested in this section is compatible with the robust counterpart reformulation discussed in Section~\ref{sect:affine_rule}.}
 \vspace{-.3em}
\section{Numerical Results}\label{sect:result}
\change{The dynamics of buildings are generally slow and can be effectively approximated using linear models, where the use of Willems' fundamental lemma is justified by real-world experiments~\cite{lian2021adaptive}. Though nonlinearity may be present, particularly the bilinearity in valve position control, there is a way to lift the nonlinear term and retain a linear analysis in the controller design~\cite{sturzenegger2015model}.} This section validates the efficacy of the proposed method using real-world I/O data collected from a building called the \textit{Polydome} located on the EPFL campus, which is a 600 $m^2$ self-standing building accommodating up to 200 people in a single lecture hall. An AERMEC RTY-04 heat pump (HP) is used to control the indoor climate. The dataset used in this study covers 40 days from December 2021 to January 2022 (i.e. the heating season) and includes indoor temperature as the output variable, the HP's electrical power consumption as the controlled input, and outdoor temperature and solar radiation as process disturbances (uncontrolled inputs) with a 15-minute sampling time. \change{Interested readers are refered to~\cite{lian2021adaptive} for more technical details.}
In the sequel, the proposed method is validated by indoor temperature control and DR service. \change{All the optimization problems are solved by \textsc{Gurobi} with Intel Core i7-1165G7 2.80 GHz processor. The solution time for different case studies are reported in the extended version. }
 \vspace{-1.8em}
\subsection{Case Study I: Temperature Consistency}
When heating is provided, the temperature consistency~\eqref{eqn:rule_temp} is enforced by the filter~\eqref{eqn:filter}. The Hankel matrices are constructed by 384 data points (i.e 4-day data for training) with $t_{init}=6$. For comparison, a parametric \change{autoregressive exogenous (ARX)} model is also considered where the physical rule is enforced by forcing the ARX weights to be positive.

As the control input is determined based on the predictor, we first run a comparison of prediction accuracy. The result is presented in Table~\ref{tab:num_case1_pred}, where different prediction horizons are considered. Even though the filtered data gives a lower prediction performance than the raw data, this performance loss results in more reasonable decisions during operation with a predictive controller. In particular, \change{consider }the following predictive control problem:
\begin{subequations}\label{eqn:num_mpc}
\begin{align}
    \min\limits_{u_{pred}} \;&\; \lVert y_{pred} - \text{ref} \rVert^2 \nonumber\\
    \text{s.t.} &\;\;  u_{pred}\in \left [0, \; 6 \, \text{kW} \right ] \nonumber\\
        & \; \; y_{pred} \; \text{by} \; \eqref{eqn:filter_ypred}\&\eqref{eqn:filter_lower_cons} \; \text{or} \; \eqref{eqn:pred_split} \nonumber
\end{align}
\end{subequations}
This controller tracks a reference temperature while considering indoor temperature constraints, and the open-loop input sequences given by different predictors are shown in Figure~\ref{fig:MPC_open_loop}. \change{The decision from the filtered-data controller maintains a maximal input before the predicted temperature reaches the reference, which is optimal regarding the turnpike property of optimal control~\cite{faulwasser2022turnpike}. While such optimal decision is also made by the parametric model, its low prediction accuracy leads to an underestimate in temperature response. This may also cause undesired chattering behaviour when the building operates around the constraint. Using two controllers defined by raw data as a comparison, their input sequences are suboptimal as their inputs oscillate between maximal input and null before raising the temperature to the reference. Note that multiple steps in open-loop input might be used in some specific applications, such as multi-building coordination. The sub-optimality observed here could deteriorate the closed-loop performance. } On top of the lack of physical consistency, we believe that these two predictors overfit, as our data is collected during the normal operation of the building, and the patterns in the I/O sequences are quite limited even though the persistent excitation condition is satisfied.


\begin{table}[h!]
\begin{center}
\caption{Comparison of the mean absolute error (MAE) over different prediction horizons} 
\begin{tabular}{  b{1cm}  b{1cm} b{1cm} b{1cm} b{1cm} b{1cm}} 
  \hline
  Prediction steps & Hours ahead & Filtered no split & Raw  split & Raw \;\; no split & Positive ARX \\ 
  \hline
  6  & 1.5 & 0.2\change{35} & 0.2\change{26} & 0.2\change{26} & 0.3\change{03}\\ 
  12  & 4 & 0.3\change{26} & 0.3\change{01} & 0.\change{299} & 0.4\change{33}\\ 
  18  & 4.5 & 0.4\change{40} & 0.3\change{92} & 0.3\change{88} & 0.5\change{89}\\ 
  \hline  
\end{tabular}
\\ \vspace{0.1ex}
{\raggedleft  ``split": horizon splitting with $n_h = t_{init}$; ``no split": otherwise\par}
\label{tab:num_case1_pred}
 \end{center}
\end{table}


\definecolor{myblue}{rgb}{0.20, 0.6, 0.78}
\definecolor{mygreen}{rgb}{0.2,0.8,0.2}
\definecolor{myred}{rgb}{0.5,0,0}
\definecolor{myorange}{rgb}{1,0.6,0.07}
\definecolor{mygrey}{rgb}{0.5,0.5,0.5}

 \begin{figure*}[htb!]
    \centering
    \begin{tikzpicture}
    \begin{groupplot}[
        legend columns=3,
        legend style={
    	font=\tiny},
        group style=
            {columns=4, horizontal sep=0.7cm}
        ]
    \nextgroupplot[
    title style={yshift=-3.2cm},
    title=(a),
    xmin= 0, xmax = 18,
    ymin= 14.5, ymax = 23,
    ytick distance = 4,
    enlargelimits=false,
    clip=true,
    grid=major,
    mark size=0.5pt,
    width=.28\linewidth,
    height=0.2\linewidth,
    ylabel = {Indoor temperature},
    xlabel={Time step $t$},
    ylabel style={at={(axis description cs:-0.1,0.5)}},
    xlabel style={at={(axis description cs:0.5,-0.1)}},
    legend columns=4,
    label style={font=\scriptsize}, 
    ticklabel style = {font=\tiny},
    legend to name=grouplegend,
    legend style={
    	font=\footnotesize, 
    	draw=none,
		at={(0.5,1.03)},
        anchor=south
    }    
    ]
    \pgfplotstableread[col sep = comma]{figs/MPC_open_loop_2.dat}{\dat}
    \addplot+ [thick,mark=*,myred,mark options={solid, myred, scale=1},forget plot] table [x={t}, y={y1}] {\dat}; 
    \addlegendimage{thick,mark=*,myred,mark options={solid, myred, scale=0.5}}   
    \addlegendentry{Output: indoor temperature}
    \addlegendimage{thick,mark=*,myblue,mark options={solid, myblue, scale=0.5}}
    \addlegendentry{Input: electrical power}  
    \addplot+ [thick,mark=none,myorange, line width=1.5pt,mark options={solid, scale=1}] table [x={t}, y expr= 22] {\dat}; 
    \addlegendentry{Output reference}   
    \addlegendimage{thick,mark=none,mygreen, line width=1.5pt,mark options={solid, myblue, scale=1}}
    \addlegendentry{Maximal input}    
    
    \nextgroupplot[
    title style={yshift=-3.2cm},
    title=(b),
    xmin= 0, xmax = 18,
    ymin= 14.5, ymax = 23,
    ytick distance = 4,
    enlargelimits=false,
    clip=true,
    grid=major,
    mark size=0.5pt,
    width=.28\linewidth,
    height=0.2\linewidth,
    ylabel = {},
    xlabel={Time step $t$},
    ylabel style={at={(axis description cs:-0.1,0.5)}},
    xlabel style={at={(axis description cs:0.5,-0.1)}},
    legend columns=2,
    label style={font=\scriptsize}, 
    ticklabel style = {font=\tiny},
    legend style={
    	font=\tiny, 
    	draw=none,
		at={(0.5,1.03)},
        anchor=south
    }    
    ]
    \pgfplotstableread[col sep = comma]{figs/MPC_open_loop_2.dat}{\dat}
    \addplot+ [thick,mark=*,myred,mark options={solid, myred, scale=1}] table [x={t}, y={y2}] {\dat}; 

    \addplot+ [thick,mark=none,myorange, line width=1.5pt,mark options={solid, myred, scale=1}] table [x={t}, y expr= 22] {\dat}; 

    \nextgroupplot[
    title style={yshift=-3.2cm},
    title=(c),
    xmin= 0, xmax = 18,
    ymin= 14.5, ymax = 23,
    ytick distance = 4,
    enlargelimits=false,
    clip=true,
    grid=major,
    mark size=0.5pt,
    width=.28\linewidth,
    height=0.2\linewidth,
    ylabel = {},
    xlabel={Time step $t$},
    ylabel style={at={(axis description cs:-0.1,0.5)}},
    xlabel style={at={(axis description cs:0.5,-0.1)}},
    legend columns=2,
    label style={font=\scriptsize}, 
    ticklabel style = {font=\tiny},
    legend style={
    	font=\tiny, 
    	draw=none,
		at={(0.5,1.03)},
        anchor=south
    }    
    ]
    \pgfplotstableread[col sep = comma]{figs/MPC_open_loop_2.dat}{\dat}
    \addplot+ [thick,mark=*,myred,mark options={solid, myred, scale=1}] table [x={t}, y={y3}] {\dat}; 
    \addplot+ [thick,mark=none,myorange, line width=1.5pt,mark options={solid, myred, scale=1}] table [x={t}, y expr= 22] {\dat};  

    \nextgroupplot[
    title style={yshift=-3.2cm},
    title=(d),
    xmin= 0, xmax = 18,
    ymin= 14.5, ymax = 23,
    ytick distance = 4,
    enlargelimits=false,
    clip=true,
    grid=major,
    mark size=0.5pt,
    width=.28\linewidth,
    height=0.2\linewidth,
    ylabel = {},
    xlabel={Time step $t$},
    ylabel style={at={(axis description cs:-0.1,0.5)}},
    xlabel style={at={(axis description cs:0.5,-0.1)}},
    legend columns=2,
    label style={font=\scriptsize}, 
    ticklabel style = {font=\tiny},
    legend style={
    	font=\tiny, 
    	draw=none,
		at={(0.5,1.03)},
        anchor=south
    }    
    ]
    \pgfplotstableread[col sep = comma]{figs/MPC_open_loop_2.dat}{\dat}
    \addplot+ [thick,mark=*,myred,mark options={solid, myred, scale=1}] table [x={t}, y={y4}] {\dat}; 
    \addplot+ [thick,mark=none,myorange,line width=1.5pt,mark options={solid, myred, scale=1}] table [x={t}, y expr= 22] {\dat};

    \end{groupplot}
    
    \begin{groupplot}[  
        legend columns=3,
        legend style={
    	font=\tiny},
        group style=
            {columns=4, horizontal sep=0.7cm}]
    \nextgroupplot[
    axis y line*=right,
    axis x line=none,
    xmin= 0, xmax = 18,
    ymin= -1, ymax = 16,  
    ytick distance = 6,
    ylabel = {},
    ylabel style={at={(axis description cs:1.1,0.5)}},      
    enlargelimits=false,
    clip=true,
    mark size=0.5pt,
    width=.28\linewidth,
    height=0.2\linewidth,
    legend columns=2,
    label style={font=\scriptsize}, 
    ticklabel style = {font=\tiny},
    legend style={
    	font=\tiny, 
    	draw=none,
		at={(0.5,1.03)},
        anchor=south
    }    
    ]  
    \pgfplotstableread[col sep = comma]{figs/MPC_open_loop_2.dat}{\dat}
    
    \addplot+ [thick,mark=none,mygreen,line width=1.5pt,mark options={solid, myblue, scale=1}] table [x={t}, y expr= 6 ] {\dat}; 
    
    \addplot+ [thick,mark=*,myblue,mark options={solid, myblue, scale=1}] table [x={t}, y={u1}] {\dat};

    \nextgroupplot[
    axis y line*=right,
    axis x line=none,
    xmin= 0, xmax = 18,
    ymin= -1, ymax = 16,  
    ytick distance = 6, 
    ylabel = {},
    ylabel style={at={(axis description cs:1.1,0.5)}},      
    enlargelimits=false,
    clip=true,
    mark size=0.5pt,
    width=.28\linewidth,
    height=0.2\linewidth,
    legend columns=2,
    label style={font=\scriptsize}, 
    ticklabel style = {font=\tiny},
    legend style={
    	font=\tiny, 
    	draw=none,
		at={(0.5,1.03)},
        anchor=south
    }    
    ]  
    \pgfplotstableread[col sep = comma]{figs/MPC_open_loop_2.dat}{\dat}
    \addplot+ [thick,mark=none,mygreen,line width=1.5pt,mark options={solid, myblue, scale=1}] table [x={t}, y expr= 6 ] {\dat};    
    \addplot+ [thick,mark=*,myblue,mark options={solid, myblue, scale=1}] table [x={t}, y={u2}] {\dat};        
    
    \nextgroupplot[
    axis y line*=right,
    axis x line=none,
    xmin= 0, xmax = 18,
    ymin= -1, ymax = 16,  
    ytick distance = 6,  
    ylabel = {},
    ylabel style={at={(axis description cs:1.1,0.5)}},      
    enlargelimits=false,
    clip=true,
    mark size=0.5pt,
    width=.28\linewidth,
    height=0.2\linewidth,
    legend columns=2,
    label style={font=\scriptsize}, 
    ticklabel style = {font=\tiny},
    legend style={
    	font=\tiny, 
    	draw=none,
		at={(0.5,1.03)},
        anchor=south
    }    
    ]  
    \pgfplotstableread[col sep = comma]{figs/MPC_open_loop_2.dat}{\dat}
     \addplot+ [thick,mark=none,mygreen,line width=1.5pt,mark options={solid, myblue, scale=1}] table [x={t}, y expr= 6 ] {\dat};   
    \addplot+ [thick,mark=*,myblue,mark options={solid, myblue, scale=1}] table [x={t}, y={u3}] {\dat};   

    \nextgroupplot[
    axis y line*=right,
    axis x line=none,
    xmin= 0, xmax = 18,
    ymin= -1, ymax = 16,  
    ytick distance = 6,  
    ylabel = {Electrical power },
    ylabel style={at={(axis description cs:1.1,0.5)}},      
    enlargelimits=false,
    clip=true,
    mark size=0.5pt,
    width=.28\linewidth,
    height=0.2\linewidth,
    legend columns=2,
    label style={font=\scriptsize}, 
    ticklabel style = {font=\tiny},
    legend style={
    	font=\tiny, 
    	draw=none,
		at={(0.5,1.03)},
        anchor=south
    }    
    ]  
    \pgfplotstableread[col sep = comma]{figs/MPC_open_loop_2.dat}{\dat}
    \addplot+ [thick,mark=none,mygreen,line width=1.5pt,mark options={solid, myblue, scale=1}] table [x={t}, y expr= 6 ] {\dat};    
    \addplot+ [thick,mark=*,myblue,mark options={solid, myblue, scale=1}] table [x={t}, y={u4}] {\dat};

    \end{groupplot}
    \path (group c2r1.north east) -- node[above]{\ref{grouplegend}} (group c3r1.north west);    
    \end{tikzpicture}
    \caption{Open-loop solution of MPC. (a)  Filtered data, no split; (b) Raw data, split ; (c) Raw data, no split; \change{(d)} Positive ARX model} 
    \label{fig:MPC_open_loop}  
    \vspace{-.61em}
 \end{figure*}
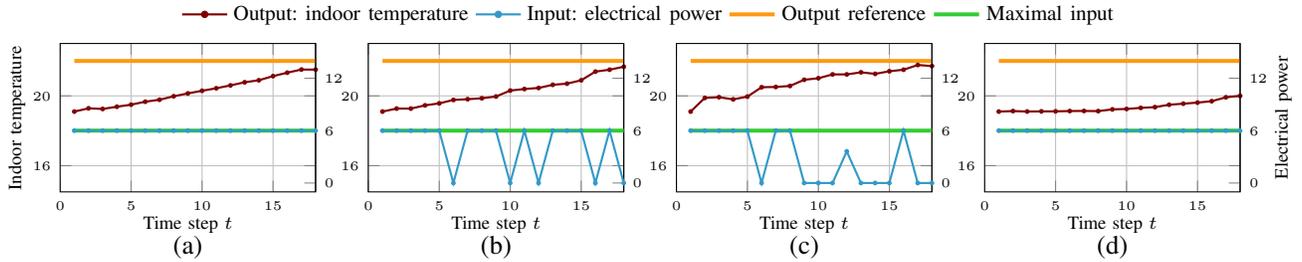





 \vspace{-1.8em}
\subsection{Case Study II: Bidding Consistency}
In this section, we consider the case where buildings are used to provide DR services and hence bidding consistency~\eqref{eqn:rule_bidding} is used. Due to a much longer prediction horizon (i.e. 24 hours), a lower sampling time, 30 minutes, is used to lower the computational cost.
The Hankel matrices are constructed by 384 data points (i.e. 8-day data) with $t_{init}=12$, and the parametric model is dropped due to the lack of convergence in its highly non-convex optimization problem.

Similar to the last part, the prediction performance is first tested on the whole dataset with different prediction horizons (see Table~\ref{tab:num_case2_pred}). In accordance with~\cite{o2022data}, splitting improves long-term prediction accuracy when we compare the results in the last two columns. However, the predictor using filtered data and splitting still gives a slightly lower prediction accuracy in comparison with the predictor generated by raw data with splitting.

These three data-driven predictors are respectively used to solve the following bidding problem:
    \begin{subequations}\label{eqn:num_DR}
\begin{align}
    \min\limits_{\gamma,P_{\text{baseline}}} \;&\; -\gamma \label{eqn:num_DR_obj}\\
    \text{s.t.} &\;\;  u_{pred,i} \in \left [0, \; 6 \, \text{kW} \right ] \nonumber\\
        &\;\;  y_{pred,i} \in \left [y_{min}, \; y_{max} \right ]\label{eqn:num_DR_ycons} \\
        & \; \; u_{pred,i} = P_{\text{baseline}} + \gamma AGC_i, i = 1,2,\dots, N_{scen}  \nonumber\\
        & \; \; y_{pred,i} \; \text{by} \; \eqref{eqn:filter_ypred}\&\eqref{eqn:filter_lower_cons} \; \text{or} \; \eqref{eqn:pred_split}\;, \nonumber 
\end{align}
\end{subequations}
where a 24-hour-ahead prediction is made within this problem. More specifically, the input flexibility margin $\gamma$ is maximized with respect to the uncertain AGC signals, whose uncertainty is handled by a scenario approach with $N_{scen}$ historical scenarios. \change{Depending on the 24-hour open-loop input decision $u_{pred}$,} $\gamma$ determines the primary remuneration from the TSO. Hence, it should be planned and sent to the TSO before the next operational day (i.e 24-hour-ahead). Interested reader are referred to~\cite{fabietti2018multi} for more technical details. To keep a compact presentation, only the data-driven predictors based on filtered/raw data with splitting are considered in this comparison. We test different comfort ranges for the indoor temperature in Table~\ref{tab:num_DR}, whose initial indoor climate and weather conditions were selected randomly from the real-world dataset.  \change{ When Problem (12) is infeasible, it is relaxed to a soft-constrained problem by relaxing (12b)  and including its violation to the cost (12a) with a large penalty}. This is done to facilitate better comparison, particularly when the temperature constraint is overly tight, such as $\left [19, 20.5\right ]$. When the constraint is set to $\left [19, 20.5\right ]$, the problem should be infeasible due to the limited power of the HVAC system \change{(i.e. $\gamma \approx 0$ in the relaxed problem)}. \change{Capturing such infeasibility is critical to avoid economic loss, and it is achieved by the problem with filtered data.} However, due to the inconsistency presented in the raw data, the problem remains feasible when the raw data is directly used. To better visualize how the physical inconsistency takes effect, we plot the control policy at different temperature constraints in Figure 2. \change{As indicated by Figure 2 (b) and Table III, a larger average heating input is applied} in the case of $y\in\left [19, 20.5\right ]$ than that in the case of $y\in\left [19, \change{22.5}\right ]$. However, it predicts a lower \change{average}
    indoor temperature, which is inconsistent with the enforced physical rule. Hence, the $\gamma$ bid based on raw data is an overestimate, and may cause indoor discomfort or economic loss in the following operational day.

\begin{table}[h!]
\begin{center}
\caption{Comparison of the MAE  over different prediction steps by three methods.} \label{tab:num_case2_pred}
\begin{tabular}{  b{1cm}  b{1cm} b{1cm} b{1cm} b{1cm} b{1cm}} 
  \hline
  Prediction steps & Hours ahead & Filtered split& Raw split &Filtered no split& Raw \;\; no split \\ 
  \hline
  12  & 6 & 0.3\change{67} & 0.3\change{44} &\change{0.367} & 0.3\change{44} \\ 
  24  & 12 & 0.4\change{96} & 0.4\change{76} &\change{0.588} & 0.4\change{94} \\ 
  36  & 18 & 0.5\change{72} & 0.5\change{09} &\change{0.739} & 0.6\change{20} \\ 
  48  & 24 & 0.6\change{08} & 0.5\change{26} &\change{0.917} & 0.7\change{80} \\ 
  \hline  
\end{tabular}
\\ \vspace{0.1ex}
{\raggedleft ``split": horizon splitting with $n_h = t_{init}$; ``no split": otherwise\par}
 \end{center}
\end{table}

\begin{table}[!htbp]
\centering
\caption{Comparison of bidding}
\begin{tabular}{*7r}
\toprule
$\left [y_{min}, \; y_{max} \right ]$ &  \multicolumn{3}{c}{Filtered data} & \multicolumn{3}{c}{Original data}\\
\midrule
{}   & $\gamma$   & $\bar{u}_{pred}$   & $\bar{y}_{pred}$  & $\gamma$   & $\bar{u}_{pred}$   & $\bar{y}_{pred}$\\
$\left [19, 20.5\right ]$ & $^{\ast}$0.00  & $^{\ast}$3.10 & $^{\ast}$20.01   & 0.94 & 4.53 & 19.88 \\
$\left [19, 21.5\right ]$ & 1.82  & 3.88 & 20.24   &  3.02 & 3.87 & 20.24 \\
$\left [19, 22.5\right ]$ & 1.94  & 4.38 & 20.45   &  3.76 & 3.81 & 20.51\\
\bottomrule
\end{tabular}
\\ \vspace{0.1ex}
{\raggedleft $\bar{u}$ and $\bar{y}$ indicate the average value \par}
{\raggedleft $^{\ast}$: from soft-constrained solution\par}
\label{tab:num_DR}
\end{table}

\definecolor{myblue}{rgb}{0.20, 0.6, 0.78}
\definecolor{mygreen}{rgb}{0.2,0.8,0.2}
\definecolor{myred}{rgb}{0.5,0,0}
\definecolor{myorange}{rgb}{1,0.6,0.07}
\definecolor{mygrey}{rgb}{0.5,0.5,0.5}

 \begin{figure}[htb!]
    \centering
    \begin{tikzpicture}
    \begin{groupplot}[
        legend columns=3,
        legend style={
    	font=\tiny},
        group style=
            {rows=2, horizontal sep=1.3cm}
        ]
    \nextgroupplot[
    title style={yshift=-3.9cm},
    title=(a),
    xmin= 0, xmax = 50,
    ymin= 14.5, ymax = 23,
    ytick distance = 2,
    enlargelimits=false,
    clip=true,
    grid=major,
    mark size=0.5pt,
    width=.9\linewidth,
    height=0.5\linewidth,
    ylabel = {Indoor temperature},
    xlabel={Time step $t$},
    ylabel style={at={(axis description cs:-0.05,0.5)}},
    xlabel style={at={(axis description cs:0.5,-0.05)}},
    legend columns=2,
    label style={font=\scriptsize}, 
    ticklabel style = {font=\tiny},
    legend style={
    	font=\scriptsize, 
    	draw=none,
		at={(0.5,1.03)},
        anchor=south
    }    
    ]
    \pgfplotstableread[col sep = comma]{figs/DR_y1_205_y2_225.dat}{\dat}
    \addplot+ [thick,mark=none,myred,mark options={solid, myred, scale=1}] table [x={t}, y={y1f}] {\dat}; 
    \addlegendentry{Output ($\left [19, 20.5\right ]$)}
    \addlegendimage{thick,mark=none,myblue,mark options={solid, myblue, scale=1}}
    \addlegendentry{Input ($\left [19, 20.5\right ]$)}  
    \addplot+ [thick,mark=none,myred, dashed, mark options={solid, myred, scale=1}] table [x={t}, y={y2f}] {\dat}; 
    \addlegendentry{Output ($\left [19, 22.5\right ]$)}   
    \addlegendimage{thick,mark=none,myblue, dashed, mark options={solid, myblue, scale=1}}
    \addlegendentry{Input ($\left [19, 22.5\right ]$)}

    \nextgroupplot[
    title style={yshift=-3.9cm},
    title=(b),
    xmin= 0, xmax = 50,
    ymin= 14.5, ymax = 23,
    enlargelimits=false,
    clip=true,
    grid=major,
    mark size=0.5pt,
    width=.9\linewidth,
    height=0.5\linewidth,
    ylabel = {Indoor temperature},
    xlabel={Time step $t$},
    ylabel style={at={(axis description cs:-0.05,0.5)}},
    xlabel style={at={(axis description cs:0.5,-0.05)}},
    legend columns=2,
    label style={font=\scriptsize}, 
    ticklabel style = {font=\tiny},
    legend style={
    	font=\footnotesize, 
    	draw=none,
		at={(0.5,1.03)},
        anchor=south
    }    
    ]
    \pgfplotstableread[col sep = comma]{figs/DR_y1_205_y2_225.dat}{\dat}
    \addplot+ [thick,mark=none,myred,mark options={solid, myred, scale=1}] table [x={t}, y={y1}] {\dat}; 
    \addlegendimage{thick,mark=none,myblue,mark options={solid, myblue, scale=1}}
    \addplot+ [thick,mark=none,myred, dashed, mark options={solid, myred, scale=1}] table [x={t}, y={y2}] {\dat}; 
    
    \end{groupplot}
    
    \begin{groupplot}[
        legend columns=3,
        legend style={
    	font=\tiny},
        group style=
            {rows=2, horizontal sep=1.3cm}
        ]
    \nextgroupplot[
    axis y line*=right,
    axis x line=none,
    xmin= 0, xmax = 50,
    ymin= -1, ymax = 16,  
    ytick distance = 6,
    ylabel = {Electrical power },
    ylabel style={at={(axis description cs:1.05,0.5)}},      
    enlargelimits=false,
    clip=true,
    mark size=0.5pt,
    width=.9\linewidth,
    height=0.5\linewidth,
    legend columns=2,
    label style={font=\scriptsize}, 
    ticklabel style = {font=\tiny},
    legend style={
    	font=\footnotesize, 
    	draw=none,
		at={(0.5,1.03)},
        anchor=south
    }    
    ]  
    \pgfplotstableread[col sep = comma]{figs/DR_y1_205_y2_225.dat}{\dat}
    \addplot+ [thick,mark=none,myblue,mark options={solid, myblue, scale=1}] table [x={t}, y={u1f}] {\dat};   
    
    \addplot+ [thick,mark=none,myblue,dashed,mark options={solid, myblue, scale=1}] table [x={t}, y={u2f} ] {\dat};        
    
    \nextgroupplot[
    axis y line*=right,
    axis x line=none,
    xmin= 0, xmax = 50,
    ymin= -1, ymax = 16,  
    ytick distance = 6,
    ylabel = {Electrical power },
    ylabel style={at={(axis description cs:1.05,0.5)}},      
    enlargelimits=false,
    clip=true,
    mark size=0.5pt,
    width=.9\linewidth,
    height=0.5\linewidth,
    legend columns=2,
    label style={font=\scriptsize}, 
    ticklabel style = {font=\tiny},
    legend style={
    	font=\tiny, 
    	draw=none,
		at={(0.5,1.03)},
        anchor=south
    }    
    ]  
    \pgfplotstableread[col sep = comma]{figs/DR_y1_205_y2_225.dat}{\dat}
    \addplot+ [thick,mark=none,myblue,mark options={solid, myblue, scale=1}] table [x={t}, y={u1}] {\dat};   
    
    \addplot+ [thick,mark=none,myblue,dashed,mark options={solid, myblue, scale=1}] table [x={t}, y={u2} ] {\dat};

    \end{groupplot}
    \end{tikzpicture}
    \caption{Solution of the demand response problem. (a)  Filtered data, split; (b) Raw data, split}
    \label{fig:DR_compare}  
 \end{figure}
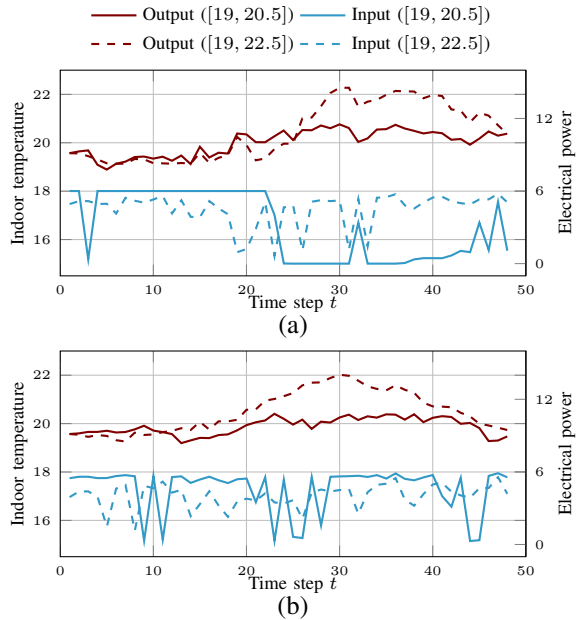

\subsection{Computation time}
In this section, we present a summary of the computation times for solving various optimization problems in Case Study I and II, as shown in Tables IV and V. It is worth noting that the predictive controller and bid problem can be solved efficiently with short computation times. Conversely, the filter requires more time, especially when using splitting, but it is only required when updating the data for Hankel matrices. Thus, considering the computing time, the proposed algorithm can be efficiently implemented online in building systems if a suitable data updating frequency is chosen.

\begin{table}[h!]
\begin{center}
\caption{Computation time of the optimization problems in Case Study I} 
\begin{tabular}{  r r r r} 
  \hline
  \shortstack[r]{Prediction \\ steps} & \shortstack[r]{Hours \\ ahead} & Filter & Controller \\ 
  \hline
  6  & 1.5  & 15.911s & 0.008s\\ 
  12  & 4  & 45.532s & 0.013s \\ 
  18  & 4.5  & 111.541s & 0.020s\\ 
  \hline  
\end{tabular}
\\ \vspace{0.1ex}
{\raggedleft  ``s": second \quad\quad\quad\quad\quad\quad \par}
\label{tab:time_case1}
 \end{center}
\end{table}

\begin{table}[h!]
\begin{center}
\caption{Computation time of the optimization problems in Case Study II} \label{tab:time_case2}
\begin{tabular}{  r r r r r r} 
  \hline
  \shortstack[r]{Prediction \\ steps} & \shortstack[r]{Hours \\ahead} & \shortstack[r]{No Split: \\filter} & \shortstack[r]{No Split: \\ bid }& \shortstack[r]{split: \\ filter} & \shortstack[r]{split: \\ bid}\\ 
  \hline
  12  & 6 & 4.59s & 0.10s & 4.59s & 0.10s \\ 
  24  & 12 & 7.02s  & 0.21s & 6.27m & 0.28s \\ 
  36  & 18 & 11.41s & 1.01s & 23.91m & 0.73s \\ 
  48  & 24 & 20.39s & 2.45s & 71.32m & 1.42s \\ 
  \hline  
\end{tabular}
\\ \vspace{0.1ex}
{\raggedleft ``s": second, ``m": minute \quad\quad \par}
 \end{center}
\end{table}

In two case studies, the sampling times are chosen according to our previous experience and experiments~\cite{lian2021adaptive,fabietti2018multi}. They were determined by practice based on the time constants of the building. 
Recently, there is research exploring the effects of the time intervals for model discretization and control sampling in building systems~\cite{huang2021simulation,gholamibozanjani2018model}. Sensitivity analysis is commonly used to determine the best choices without theoretical guarantees. Here gives an example of analysis on the prediction error and computation time  for 8-day data in Figure~\ref{fig:sample_time}. 
As the sampling time increases, the 6-hour prediction error does not change much but the filtering computation time decreases a lot. In fact, it is an interesting future direction to perform  more analysis, such as the one for closed-loop control performance by some simulation software.

\definecolor{myblue}{rgb}{0.20, 0.6, 0.78}
\definecolor{mygreen}{rgb}{0.2,0.8,0.2}
\definecolor{myred}{rgb}{0.5,0,0}
\definecolor{myorange}{rgb}{1,0.6,0.07}
\definecolor{mygrey}{rgb}{0.5,0.5,0.5}

 \begin{figure}[htb!]
    \centering
    \begin{tikzpicture}
    \begin{groupplot}[
        legend columns=3,
        legend style={
    	font=\tiny},
        group style=
            {rows=2, horizontal sep=1.3cm}
        ]
    \nextgroupplot[
    title style={yshift=-3.9cm},
    title=(a),
    xmin= 13, xmax = 62,
    ymin= 0, ymax = 0.6,
    xtick distance = 15,    
    ytick distance = 0.2,
    enlargelimits=false,
    clip=true,
    grid=major,
    mark size=0.5pt,
    width=.9\linewidth,
    height=0.5\linewidth,
    ylabel = {Prediction error},
    xlabel={Sampling time [minute] },
    ylabel style={at={(axis description cs:-0.06,0.5)}},
    xlabel style={at={(axis description cs:0.5,-0.05)}},
    legend columns=2,
    label style={font=\scriptsize}, 
    ticklabel style = {font=\tiny},
    legend style={
    	font=\scriptsize, 
    	draw=none,
		at={(0.5,1.03)},
        anchor=south
    }    
    ]

    \addplot+ [thick,mark=square,myred,mark options={solid, myred, scale=3}] table [x=x,y=y,row sep=crcr] { x      y \\
     15      0.416 \\
     30     0.367 \\
     45     0.362 \\
     60     0.368\\};  
    \addlegendentry{Prediction error}
    
    \addlegendimage{thick,mark=square,myblue,mark options={ solid, myblue, scale=3}}
    \addlegendentry{Computation time}

    \nextgroupplot[
    title style={yshift=-3.9cm},
    title=(b),
    xmin= 13, xmax = 62,
    ymin= 0, ymax = 45,
    xtick distance = 15,    
    ytick distance = 15,
    enlargelimits=false,
    clip=true,
    grid=major,
    mark size=0.5pt,
    width=.9\linewidth,
    height=0.5\linewidth,
    ylabel = {Compuation time [second]},
    xlabel={Sampling time [minute] },
    ylabel style={at={(axis description cs:-0.06,0.5)}},
    xlabel style={at={(axis description cs:0.5,-0.05)}},
    legend columns=2,
    label style={font=\scriptsize}, 
    ticklabel style = {font=\tiny},
    legend style={
    	font=\footnotesize, 
    	draw=none,
		at={(0.5,1.03)},
        anchor=south
    }    
    ]

    \addplot+ [thick,mark=square,myblue,mark options={ solid, myblue, scale=3}] table [x=x,y=y,row sep=crcr] { x      y \\
     15      36.683 \\
     30     4.833 \\
     45     1.506 \\
     60     0.720\\};

    \end{groupplot}

    \end{tikzpicture}
    \caption{Comparison of different choices of sampling times for 6-hour ahead bidding filtering (a)  Prediction error ; (b) Computation time}
    \label{fig:sample_time}  
 \end{figure}
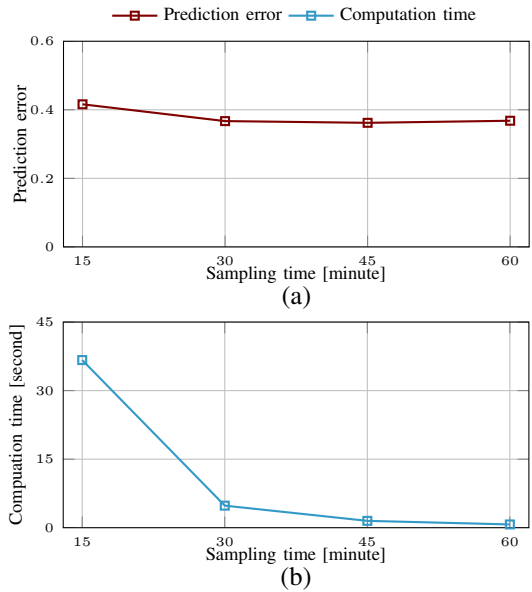
    
\section{Conclusions}\label{sect:conclusion}
In this paper, a physics-based filter was proposed to enhance data-driven predictors. The scheme enforces a priori physical rules, improving decision-making reliability.



\bibliographystyle{ieeetr}
\bibliography{ref.bib}

\end{document}